\documentclass{article}
\usepackage{epsfig}

\newcommand{\scost}{S_{\mbox{cost}}}

\title{Efficiencies and optimization of HMC algorithms in pure gauge theory}

\author{Bernd Gehrmann and Ulli Wolff\\[5mm]
Institut f\"ur Physik, Humboldt-Universit\"at zu Berlin,\\
Invalidenstr. 110, 10115 Berlin, Germany}

\begin{document}

\parindent=0cm
\addtolength{\topmargin}{-2\baselineskip}
\addtolength{\textheight}{4\baselineskip}

\maketitle

\begin{abstract}
As a prerequisite to dynamical fermion simulations a detailed study of
optimal parameters and scaling behavior is conducted for the
quenched Schr\"odinger functional at fixed renormalized coupling. We compare
standard hybrid overrelaxation techniques with local and global hybrid 
Monte Carlo. Our efficiency measure is designed to be directly
relevant for the strong coupling constant as used by the ALPHA
collaboration.
\end{abstract}

The measurements are performed 
in the framework of the Schr\"odinger
Functional, i.e.
\begin{itemize}
  \item finite lattice with extent $L^3\times T$ (here $T=L$),
  \item periodic boundary conditions in the spatial directions,
  \item fixed boundary fields $C$ and $C'$ at $x_0=0$ and $x_0=T$,
        parametrized by a dimensionless parameter $\eta$
        \cite{9309005},
\end{itemize}

\begin{equation}
  {\mathcal Z}[C,C'] = e^{-\Gamma} = \int [dU] e^{-S[U]}.
\end{equation}

We consider a pure SU(3) gauge action,
\begin{equation}
  S[U] = \beta \sum_p w(p) 
         \left( 1- \frac 1 3 \mbox{Re} \mbox{Tr} U_p \right).
\end{equation}

A renormalized coupling can be extracted from the response
of the free energy to the induced colour-electric background
field. There is some freedom in the precise definition of the
coupling. We choose one which has turned out to be practical
in the ALPHA simulations. This coupling is obtained as an 
expectation value
\begin{equation}
  \frac{1}{\bar g^2} = 
    k^{-1} \frac{\partial\Gamma}{\partial\eta} =
    k^{-1} \left\langle \frac{\partial S}{\partial\eta} \right\rangle.
\end{equation}

In this finite volume scheme, the running coupling $\bar g$ 
is a function of the box size $L$. We perform our measurements
at constant physics, i.~e. with $\beta$ tuned such that the
coupling of systems with different $L/a$ match, where $a$
is the lattice spacing.

\begin{center}
  \begin{tabular}{r|r|r}
    L/a & \multicolumn{1}{c|}{$\mathbf\beta$} 
    & \multicolumn{1}{c}{$\mathbf{\bar g}^{-2}$} \\
    \hline
    4 & 6.7796 & 0.476 \\
    6 & 7.1214 & 0.476 \\
    8 & 7.3632 & 0.476 \\
    10 & 7.5525 & 0.476 \\
  \end{tabular}
\end{center}
  
We consider the following algorithms:

\paragraph*{Heatbath/Overrelaxation (HOR).} 

Each update is composed of one heatbath sweep followed by
$N_{or}$ overrelaxation sweeps.

\begin{itemize}
\item Overrelaxation: microcanonical reflections in three SU(2) 
      subgroups.
\item Heatbath: Cabbibo-Marinari method with random matrices from
      SU(2) subgroups. SU(2) matrices are generated with the
      Fabricius-Haan algorithm.
\end{itemize}

A sweep consists of loops over $x$ and $\mu$. The order of these
loops turned out to have a significant influence on the 
autocorrelation times (with the inner loop over $x$ being in
advantage).

\paragraph*{Hybrid Monte Carlo (HMC).} 

The HMC is a member of the family of algorithms which are based
on classical dynamics. To this end, one considers
a Hamiltonian

\begin{equation}
  H[P,U] = \frac{1}{2} \sum_{x,\mu} P_{x,\mu}^2 + S[U]
\end{equation}

with momenta $P_{x\mu}$ conjugate to the link variables.
In each update step, momenta are generated with a Gaussian 
distribution and a reversible discretized trajectory is
computed. Discretization errors are corrected by an acceptance 
step.

Parameters which can be optimized are the trajectory length 
$\tau$ and the step size $\delta\tau$.

\paragraph*{Local Hybrid Monte Carlo (LHMC).} 

We also consider a local version of the HMC algorithm proposed 
in \cite{Rossi}. Here one applies the same procedure as for the 
global HMC algorithm, but for one link while keeping all others 
fixed. This has some advantages compared to the global HMC:

\begin{itemize}
\item The difference in the action accumulated on a trajectory
      does not contain a volume factor. As a consequence, much
      greater step sizes sizes are possible without getting
      poor acceptance rates.
\item The staples for a link -- which are involved in the computation
      of the force -- have to be computed only once per trajectory,
      so additional steps on the trajectory are cheap. This point
      has turned out to be of minor importance, as the optimal
      parameter set has $\tau/\delta\tau \approx$ 2 -- 3. 
\end{itemize}

An advantage compared to the HOR algorithm is that $\Delta S$ is 
only needed for infinitesimal $\Delta U$. This makes no practical
difference for the standard Wilson action considered here, but it 
is relevant for more complicated actions where terms quadratic
in $U$ appear.

The LHMC algorithm cannot be applied to QCD with pseudofermions
in a straightforward way, because the computation of the force
would then require the inversion of a fermion matrix in each
local step. However, for a simulation of a bermion theory with 
clover term, which we are preparing, it is an attractive candidate 
for the update of the gauge field \cite{Bermions}. 
  
We mention here that for a Wilson action, the leapfrog algorithm
for generating a candidate configuration can be replaced by an
exact integration of the equations of motion \cite{Kennedy}.
The method used there makes use of the fact that in the case
where subgroups of SU(3) are updated separately, the action
takes the form of the energy of a "pendulum".

We want to represent efficiency measures in a way that
allows a machine-independent comparison of algorithms.
Thus, we define a measure $\scost$ such that in order to
compute ${\bar g}^2$ at $1\%$ accuracy, the equivalent 
in complexity of $\scost$ computations of all staples is 
required.

The optimization of the three algorithms used in this study
is shown in figures 1 - 5.

In the following we investigate how the different algorithms
scale with $L/a$. The variance of the coupling between $L/a=4$ 
and $L/a=10$ behaves approximately like $(L/a)^{1.4}$. Thus,
we expect a behaviour of the cost measure  
$\scost \propto (L/a)^{1.4+z}$. In figure~\ref{fig:scale}
we have plotted $\scost$ with optimized parameters for each 
algorithm against $L/a$. The dotted lines correspond to exponents
$z=1$, $z=2$ resp. 

From our data at $L/a=8$ we conclude that typical ratios in
$\scost$ for optimized parameters are 1\,:\,3\,:\,26 for
HOR : LHMC : HMC. This illustrates the cost of HMC even
before dynamical quarks are included. A similar performance
ratio for HMC and HOR was concluded in \cite{Gupta}, which
recently came to our attention.

We thank DESY for allocating computer time to this project and
the DFG under GK 271 for financial support.

\begin{figure}[hp]
\begin{center}
\epsfig{file=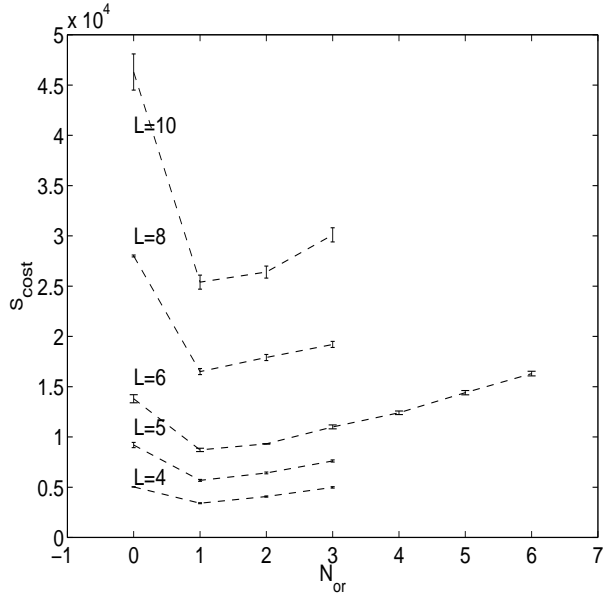,height=8cm,width=8cm}
\vspace{-0.3cm}
\caption{\label{fig:hor}
\sl $\scost$ for the HOR algorithm}
\end{center}
\end{figure}

\begin{figure}[hp]
\begin{center}
\epsfig{file=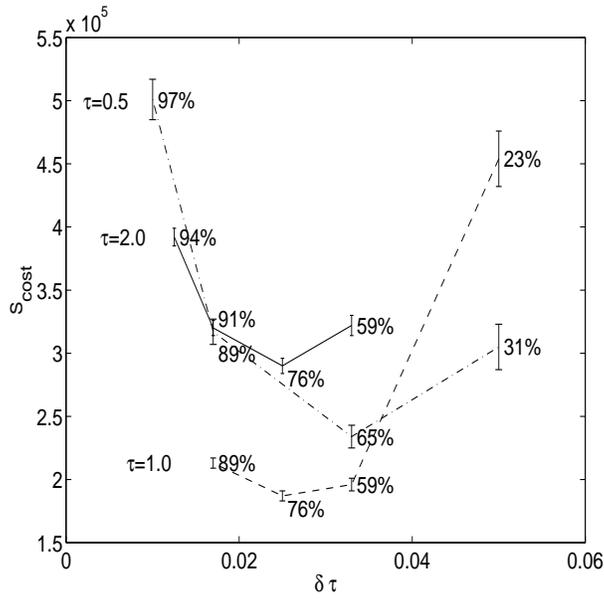,height=8cm,width=8cm} 
\vspace{-0.3cm}
\caption{\label{fig:hmc6}
\sl $\scost$ for global Hybrid Monte Carlo at $L/a=6$.
For each data point, the corresponding acceptance is shown.}
\end{center}
\end{figure}

\begin{figure}[hp]
\begin{center}
\epsfig{file=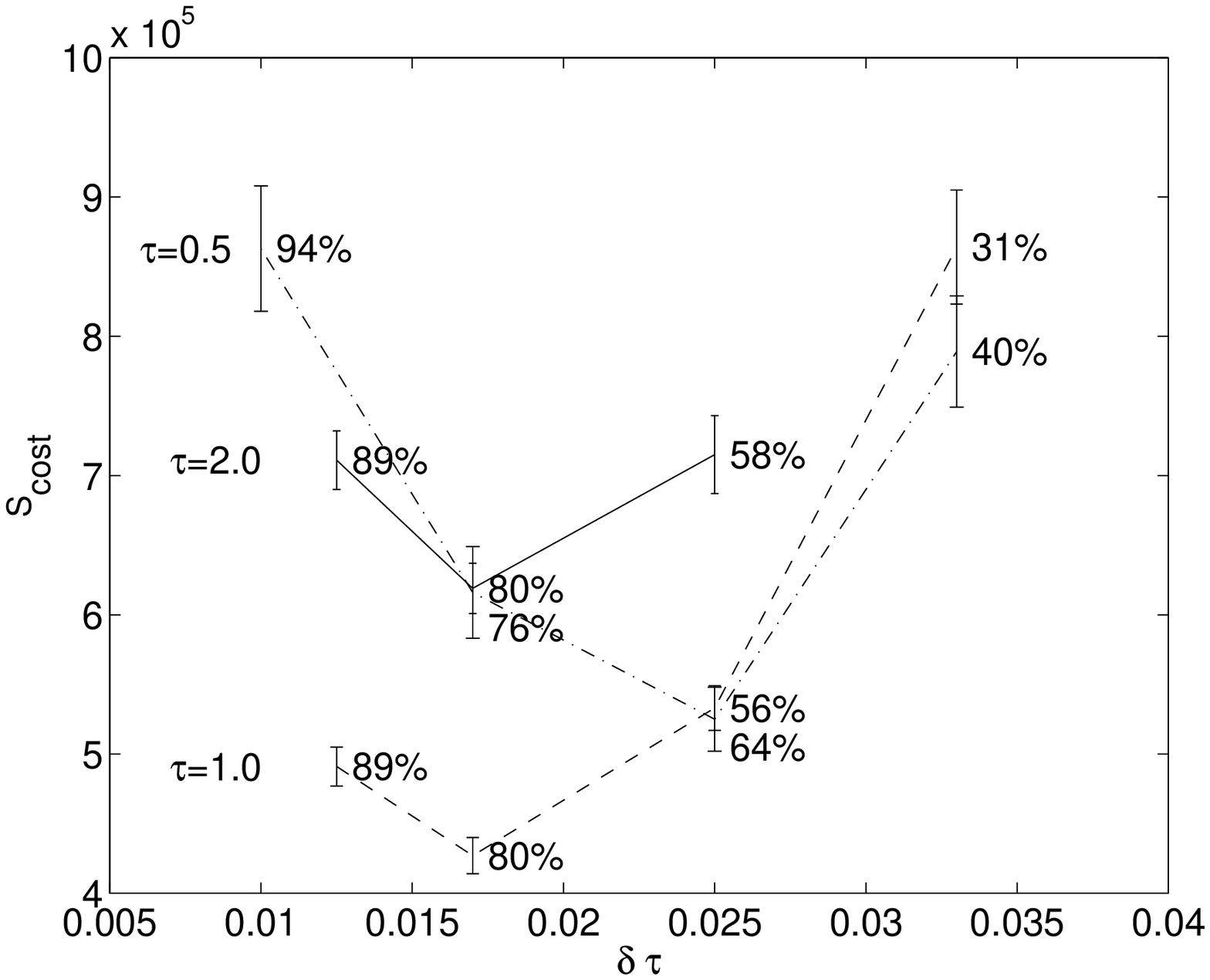,height=8cm,width=8cm} \\
\vspace{-0.3cm}
\caption{\label{fig:hmc8}
\sl $\scost$ for global Hybrid Monte Carlo at $L/a=8$.}
\end{center}
\end{figure}

\begin{figure}[hp]
\begin{center}
\epsfig{file=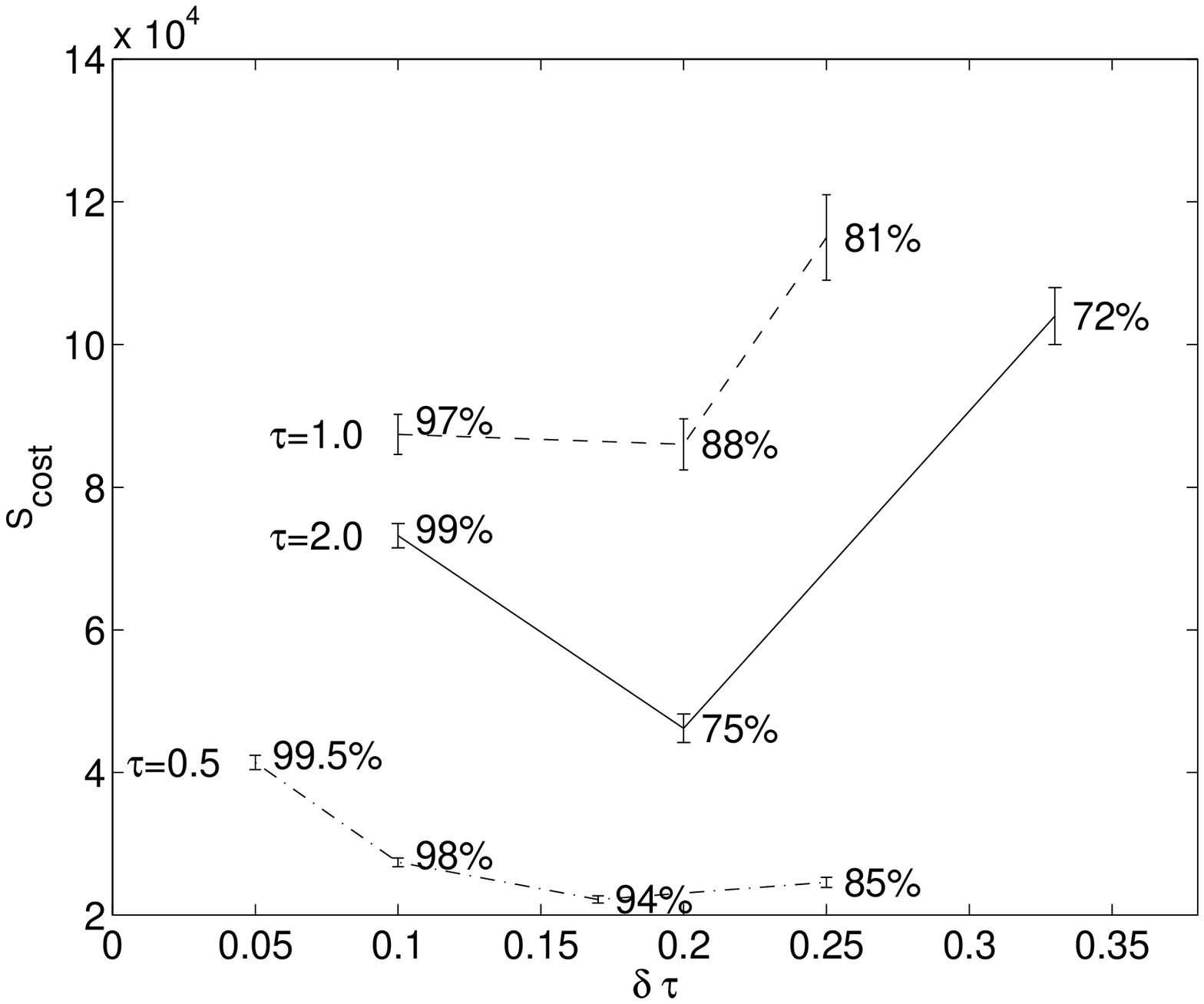,height=8cm,width=8cm}
\vspace{-0.3cm}
\caption{\label{fig:lhmc6}
\sl $\scost$ for Local Hybrid Monte Carlo at $L/a=6$.}
\end{center}
\end{figure}

\begin{figure}[hp]
\begin{center}
\epsfig{file=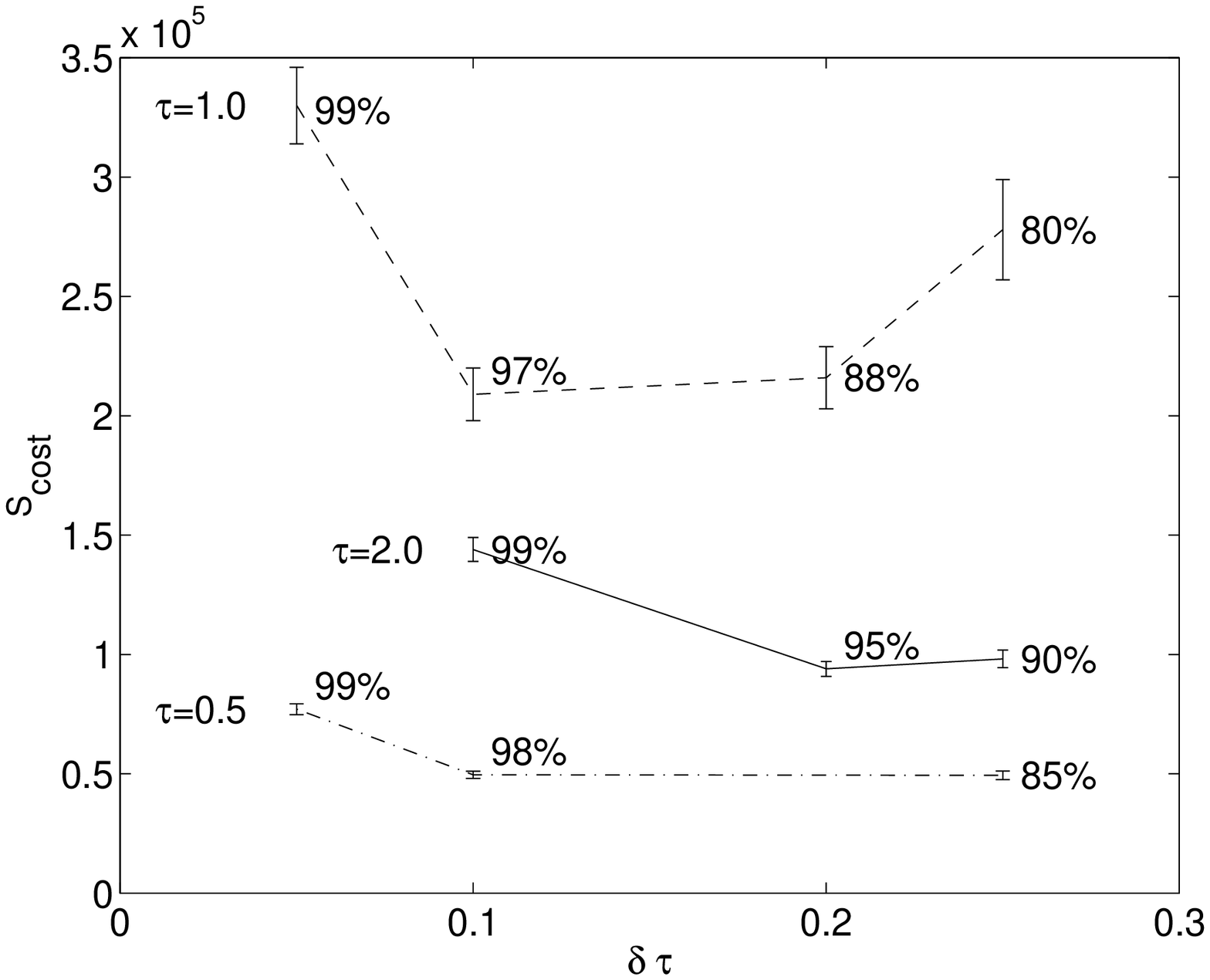,height=8cm,width=8cm} \\
\vspace{-0.3cm}
\caption{\label{fig:lhmc8}
\sl $\scost$ for Local Hybrid Monte Carlo at $L/a=8$.}
\end{center}
\end{figure}

\begin{figure}[hp]
\begin{center}
\epsfig{file=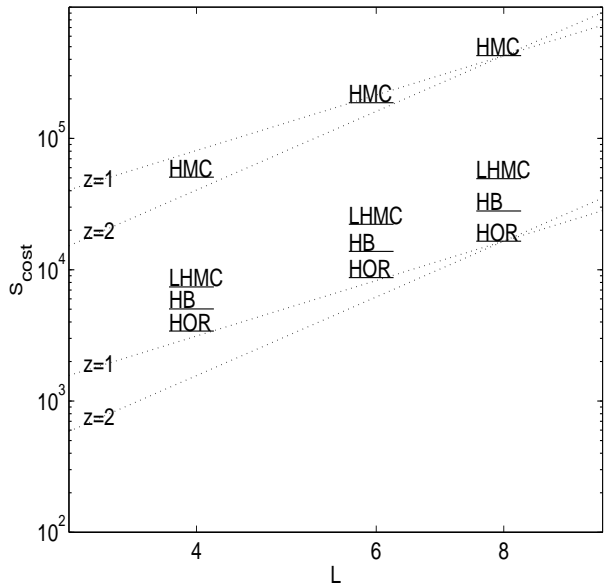,height=8cm,width=8cm}
\vspace{-0.3cm}
\caption{\label{fig:scale}
\sl Optimal $\scost$ for the different algorithms in a
double-logarithmic representation. For comparison, dotted
lines which correspond to exponents $z=1,2$ are displayed.
The points denoted with HB correspond to a pure heatbath.}
\end{center}
\end{figure}

\end{document}